\documentclass[12pt,a4paper,notitlepage]{article}%
\usepackage{epsfig}
\usepackage{amsmath}
\usepackage{graphicx}%
\usepackage{amsfonts}%
\usepackage{amssymb}
\begin{document}

\author{Micha\l\ Prasza\l owicz\thanks{\texttt{michal@if.uj.edu.pl}}\\
{\small {M. Smoluchowski Institute of Physics,} } \\
{\small {Jagellonian University}, } \\
{\small {Reymonta 4, 30-049 Krak\'{o}w, Poland}}}
\title{\flushright{\small TPJU-5/2003}  \\~~~\\
\centerline{Pentaquark in the Skyrme Model}  }
\maketitle

\begin{abstract}
The mass of the newly discovered pentaquark is calculated within the framework
of the SU(3) Skyrme Model. Various estimates based on the \emph{model
independent} approach are compared with the model results and with the Chiral
Quark Model. Our discussion shows that $\Theta^{+}$ is light with the mass of
the order 1.5 GeV.

\end{abstract}

\begin{flushleft}
{\bf
1. Introduction}
\end{flushleft}

It seems that the new exotic spin $1/2$ baryon of strangeness $+1$,
$\Theta^{+}$, is now well established experimentally \cite{nakano}%
,\cite{Diana},\cite{CLAS},\cite{SAPHIR}. After the first report by Nakano,
three other experiments have already confirmed the existence of $\Theta^{+}$
with mass $M_{\Theta^{+}}\simeq1540$ MeV and very small width $\Gamma
_{\Theta^{+}}<25$ MeV. It is interesting to note that such a state was
predicted in the late 80-ies within the framework of the Skyrme Model. Indeed,
Manohar \cite{man} and Chemtob in Ref.\cite{Chem} mentioned that the SU(3)
Skyrme Model \cite{Gua},\cite{NMP},\cite{JW} (SM) not only reproduced the
spectrum of the lowest baryon multiplets $8$ and $10$, but also predicted the
new, exotic states belonging to the higher SU(3) representations like
$\overline{10}$ or $27$. Such states were also seen in the KN scattering phase
shifts calculated within SM in Refs.\cite{MatKar},\cite{KarMog}. In 1987 in
Ref.\cite{Mogtalk} we presented the first estimate of the mass of the lightest
member of $\overline{10}$ -- today's $\Theta^{+}$ -- with the result
$M_{\Theta^{+}}\simeq1530$ MeV; in a striking agreement with the present
experimental observations. Since the details of this estimate were never
published, we think that it is worthwhile to recall shortly how this result
was actually obtained.

The real boost in this field was due to Ref.\cite{DPP} published in 1997,
where not only the mass of $\Theta^{+}$ but also its width was calculated
within the framework of the Chiral Quark Model ($\chi$QM) \cite{Drev}. In fact
both SM and $\chi$QM have similar group theoretical structure and the main
problem in both models consisted in estimating the average mass of the exotic
antidecuplet, $M_{\overline{10}}$. The splittings within $\overline{10}$ are
to a good approximation in the first order in the symmetry breaker, $m_{s}$,
identical as in the usual decuplet, \emph{i.e.} proportional to the
hypercharge $Y$, with proportionality coefficient being of the order of $150$
MeV. The problem of fixing $M_{\overline{10}}$ was solved in Ref.\cite{DPP} by
the assumption that the nucleon resonance N$^{\ast}(1710)$ belongs to
$\overline{10}$. An immediate consequence of this assignment was that
$M_{\Theta^{+}}\simeq M_{N^{\ast}}-150=1560$ MeV. A more refined analysis of
the antidecuplet splittings pushed the $\Theta^{+}$ mass further down to the
value of 1530 MeV \cite{DPP}.

In our old estimate of the $\Theta^{+}$ mass \cite{Mogtalk} we used the second
order perturbation theory in the strange quark mass $m_{s}$. This allowed us
to estimate the strange moment of inertia of the rotating soliton which is
responsible for the $\overline{10}-8$ mass splitting. Therefore our analysis
did not rely on any physical assumption assigning some existing nucleon
resonance to $\overline{10}$. The price, however, was that the mass of
$\Theta^{+}$ depended rather strongly on the value of $m_{s}$ and/or pion
nucleon $\Sigma_{\pi N}$ term.

\begin{flushleft}%
{\bf 
2. Baryons in the chiral models}
\end{flushleft}

Chiral models like $\chi$QM or SM are closely related to QCD \cite{Drev}. One
can formally \emph{integrate out} gluons from the QCD Lagrangian, then the
resulting nonlocal quark theory would necessarily respect chiral symmetry. One
can approximate this complicated theory by a simple, chirally symmetric,
quartic quark interactions. A model with such properties was formulated by
Nambu and Jona-Lasinio \cite{NJL}, however in a different context. NJL model
exhibits an important phenomenon: spontaneous chiral symmetry breaking. As a
result massless quark-antiquark bound states emerge -- pions, kaons and $\eta$
($\varphi_{a}$ in short). Formally, one can now reexpress the interaction part
of the NJL Lagrangian introducing auxiliary, composite meson fields. This is
the Chiral Quark Model used in Ref.\cite{DPP} and derived from QCD within the
instanon model of the QCD vacuum \cite{instvac}.

Notice the absence of the explicit meson kinetic energy and meson
self-interaction terms in $\chi$QM: they arise only from the quark loops when
one integrates out the quark fields. The resulting Lagrangian can be organized
in terms of a number of derivatives of the meson fields (gradient expansion)
\cite{grad}. A simple Ansatz for such a Lagrangian density was proposed by
Skyrme 40 years ago \cite{Skyrme}:%
\begin{equation}
\mathcal{L}=\frac{F_{\pi}^{2}}{16}\operatorname*{Tr}\left(  \partial_{\mu
}U^{\dagger}\partial^{\mu}U\right)  +\frac{1}{32e^{2}}\operatorname*{Tr}%
\left(  \left[  \partial_{\mu}U\,U^{\dagger},\partial_{\nu}U\,U^{\dagger
}\right]  ^{2}\right)  +\Gamma_{\text{WZ}}.\label{Lagr}%
\end{equation}
Here $U=\exp\left(  i2\varphi_{a}\lambda_{a}/F_{\pi}\right)  $, $F_{\pi}=186$
MeV, $e$ denotes the Skyrme parameter of the order $4-5$ and $\Gamma
_{\text{WZ}}$ stands for the Wess-Zumiono term \cite{WZ},\cite{Witten}.

Clearly, both $\chi$QM and SM are devised to describe meson physics at low
energies. However, as realized by Skyrme \cite{Skyrme} and later by Witten
\cite{Witten} and collaborators \cite{ANW}, there exists a solitonic solution
-- a nontrivial \emph{classical} configuration of the meson fields, taken in
the form of the \emph{hedgehog} Ansatz:%
\begin{equation}
U_{0}=\left[
\begin{array}
[c]{cc}%
e^{i\vec{n}\cdot\vec{\tau}\,P(r)} & 0\\
0 & 1
\end{array}
\right] \label{hedgehog}%
\end{equation}
-- which can be interpreted as a baryon. Further quantization of the
rotational zero modes (both in the configuration space and in the flavor
space), \emph{i.e.} rotations of the meson fields $U=AU_{0}A^{\dagger}$
parameterized by a time dependent SU(3) matrix $A(t)$, provides quantum
numbers corresponding to the different baryonic states \cite{Gua}%
,\cite{NMP},\cite{JW}. In the chiral symmetry limit the effective baryonic
Hamiltonian takes the following form:
\begin{equation}
H_{0}=M_{\text{cl}}+\frac{1}{2I_{1}}S(S+1)+\frac{1}{2I_{2}}\left(
C_{2}(\mathcal{R})-S(S+1)-\frac{N_{{c}}^{2}}{12}\right)  .\label{H0}%
\end{equation}
Here $S$ denotes baryon spin, $C_{2}(\mathcal{R})$ the Casimir operator for
the SU(3) representation $\mathcal{R}=(p,q)$:
\begin{equation}
C_{2}(\mathcal{R})=\frac{1}{3}\left(  p^{2}+q^{2}+pq+3(p+q)\right)  .
\end{equation}
The classical soliton mass $M_{\text{cl}}[P]$, and two moments of inertia
$I_{1,2}[P]$ are functionals of the solitonic solution (\ref{hedgehog}) and
can be numerically calculated within the model considered. However, as already
observed in Ref.\cite{ANW} in the SU(2) version of the Skyrme Model and
further confirmed in the SU(3) case \cite{MPSU3}, the classical soliton mass
is by far too large for the realistic set of parameters \cite{Weigel}.
Therefore we adopt here a ''model-independent'' approach of Ref.\cite{AN}
where $M_{\text{cl}}$ and $I_{1,2}$ are treated as free parameters. This
approach may be justified by the large $N_{c}$ argument. Constants
$M_{\text{cl}}$, $I_{1,2}\sim N_{c}$, hence the classical mass is of the order
$N_{c}$ while the splittings are of the order $1/N_{c}$. One can easily argue
that there are $\mathcal{O}(1)$ negative corrections to $M_{\text{cl}}$ which
are in missing the present approach. Indeed, in some cases the so called
Casimir energy was calculated for the solitonic solutions and was proven to be
negative \cite{Cas}.

Hamiltonian (\ref{H0}) has to be supplemented by a constraint which says that
the allowed SU(3) representations $\mathcal{R}=(p,q)$ have to contain states
with hypercharge $Y=1$ and the isospin of $Y=1$ states is interpreted as
(minus) baryon spin. The constraint $Y=1$, which follows from the Wess-Zumino
term, selects the representations of triality zero \cite{Gua},\cite{Chem}%
,\cite{Mogtalk}:
\begin{equation}
8,\,10,\,\overline{10},\,27,\,35,\,\overline{35},\,64,\,\ldots\label{eq:reps}%
\end{equation}
for $N_{c}=3$. The success of the model is the prediction that the lowest
baryonic states belong to the octet and decuplet representations of SU(3). The
splitting between exotic ($\overline{10},\,27$, \emph{etc.}) and nonexotic
representations depends on the strange moment of inertia $I_{2}$, whereas the
splitting between ''standard'' representations depends only on $I_{1}$.

The wave function of baryon $B=Y,I,I_{3}$ belonging to representation
$\mathcal{R}=(p,q)$ and of spin $S$ is defined as \cite{Blotz}:
\begin{equation}
\left|  R,B,S\right\rangle =\sqrt{R}\,(-)^{Q^{\prime}}D_{YII_{3}\;\;Y^{\prime
}JJ_{3}}^{(\mathcal{R})\,\ast}(A)\label{wf}%
\end{equation}
where $Y^{\prime}=N_{\mathrm{c}}/3$ and $S=S,S_{3}$ denotes spin and
$Q^{\prime}$ is charge of the state $(Y^{\prime}JJ_{3})$. $D_{ab}%
^{(\mathcal{R})}$ are the SU(3) Wigner matrices in representation
$\mathcal{R}$ \cite{DeSw}. The quantization results in the relation: $J=S$ and
$J_{3}=-S_{3}$. $R$ denotes representation and its size as well:
\begin{equation}
R=\dim(\mathcal{R})=\frac{1}{2}(p+1)(q+1)(p+q+2).
\end{equation}
Note that $A$ is an SU(3) rotation matrix and therefore baryonic wave
functions ''live'' in the SU$_{\text{flavor}}$(3)$\otimes$SU$_{\text{spin}}%
$(3) space, where the second group is constrained to the SU(2) subgroup
corresponding to spin.

\begin{flushleft}%
{\bf
3. Symmetry breaking and the mass formulae}%

\end{flushleft}

Hamiltonian (\ref{H0}) and the wave functions (\ref{wf}) are identical in both
Chiral Quark Model and Skyrme Model (with an obvious difference as far as
model formulae for $M_{\text{cl}}$ and $I_{1,2}$ are concerned). In the
leading order in the symmetry breaking parameter $m_{s}$ (strange quark mass)
and in the leading order of the $N_{c}$ expansion, the symmetry breaking
Hamiltonian looks also identically in both models:%
\begin{equation}
H^{\prime}=m_{\text{cl}}-\alpha D_{88}^{(8)}(A)=m_{s}\sigma-m_{s}\sigma
D_{88}^{(8)}(A).\label{Hprim}%
\end{equation}
Indices $a,b=8$ mean that $Y=0,I=0,\,I_{3}=0$. Here $\sigma$ is related to the
pion-nucleon sigma term:%
\begin{equation}
\sigma=\frac{1}{3}\frac{\Sigma_{\pi N}}{\overline{m}}%
\end{equation}
$\overline{m}$ being the average mass of the up and down quarks. The classical
mass $m_{\text{cl}}=m_{s}\sigma$ is a constant which will be in what follows
ignored, or more precisely included in the soliton mass $M_{\text{cl}}$ in
Eq.(\ref{H0}): $M_{\text{cl}}\rightarrow M_{\text{cl}}+m_{\text{cl}}.$

The symmetry breaking Hamiltonian $D_{88}^{(8)}$ is not reach enough to
reproduce the mass splittings in the octet \cite{MPSU3}. It has become evident
that in order to fit the mass splittings one has go beyond the leading order
either in $N_{c}$ or in $m_{s}$. In 1984 Guadagnini \cite{Gua} introduced an
additional term proportional to hypercharge, $\beta Y$, whose coefficient
$\beta$ is of the order of $\mathcal{O}(N_{c}^{0})$ and linear in $m_{s}$. In
the $\chi$QM still another nonleading term is present \cite{Blotz}, so that
the perturbation Hamiltonian takes the following form:%
\begin{equation}
H_{\chi\text{QM}}^{\prime}=-\alpha^{\prime}D_{88}^{(8)}+\beta Y+\frac{\gamma
}{\sqrt{3}}\sum_{i=1}^{3}D_{8i}^{(8)}S_{i}\label{HbrQM}%
\end{equation}
where $\alpha^{\prime}=\alpha+\tilde{\alpha}$, with $\alpha\sim\mathcal{O}%
(N_{c})$ and $\tilde{\alpha},\beta,\gamma\sim\mathcal{O}(N_{c}^{0})$ (of
course all the constants in (\ref{HbrQM}) are linear in $m_{s}$). Since the
matrix elements of $H_{\chi\text{QM}}^{\prime}$ between octet and decuplet
wave functions (\ref{wf}) depend only upon two linear combinations of the
three constants $\alpha^{\prime}$, $\beta$ and $\gamma$, the total number of
parameters is $4$; the two others being $M_{\text{cl}}-\frac{3}{4I_{2}}$ and
$\frac{1}{I_{1}}$, or equivalently $M_{8}$ and $M_{10}$ -- average octet and
decuplet masses. The above discussion shows that there is no way to extract
$I_{2}$ and, say, $\gamma$ from the spectra of the ''standard'' baryons. The
quality of the fit to the mass spectra is very good and there exists a new (as
compared to Gell-Mann--Okubo mass relations) sum rule known as the Guadagnini
relation \cite{Gua}:
\begin{equation}
8(\text{N}+\Xi^{\ast})+3\Sigma=11\Lambda+8\Sigma^{\ast}.\label{Guad}%
\end{equation}

Another point of view was advocated by Yabu and Ando \cite{YA} who in 1988
proposed to diagonalize exactly breaking Hamiltonian (\ref{Hprim}). By
analytical methods developed in Ref.\cite{YA} they reduced the problem to the
harmonic oscillator with the frequency proportional to $m_{s}$. They obtained
a very good fit to the data, although their approach was criticized because at
the level $\mathcal{O}(m_{s}^{2},m_{s}^{3},\ldots)$ there are in principle
other possible symmetry breaking contributions to $H^{\prime}$ which they
neglected. However, the same objection can be raised against $H_{\chi
\text{QM}}^{\prime}$ and $N_{c}$ power counting, with an explicit example
discussed for instance in Ref.\cite{MPgA}. Guided by the Yabu-Ando method we
have observed in \cite{Mogtalk} that already the second order perturbation
theory in $m_{s}$ gives a good approximation to the fully diagonalized
Hamiltonian. Moreover, as we will show below, the second order perturbation
theory gives us a possibility to constrain the parameter $I_{2}$ responsible
for the center of mass of the exotic antidecuplet.

Let us observe that the symmetry breaking Hamiltonian%
\begin{equation}
H_{\text{SM}}^{\prime}=-\alpha D_{88}^{(8)}(A)\label{HSM}%
\end{equation}
does not change the quantum numbers of the state on which it acts; it can only
change the representation:%
\begin{equation}
\delta_{B(R)}^{R^{\prime}}=\left\langle R^{\prime},B,S\right|  D_{88}%
^{(8)}\left|  R,B,S\right\rangle .
\end{equation}
Therefore the first order correction to the baryon mass reads%
\begin{equation}
M_{B(R)}^{(1)}=-\alpha\delta_{B(R)}^{R},\label{1st}%
\end{equation}
whereas the second order correction takes the following form%
\begin{equation}
M_{B(R)}^{(2)}=-2I_{2}\alpha^{2}\sum\limits_{R^{\prime}\neq R}\frac{(\delta
_{B(R)}^{R^{\prime}})^{2}}{\Delta_{R^{\prime}R}}\label{2nd}%
\end{equation}
where%
\begin{equation}
\Delta_{R^{\prime}R}=C_{2}(R^{\prime})-C_{2}(R).\label{Deldef}%
\end{equation}
Let us give for completeness the values $\Delta$'s:%
\[
\Delta_{\overline{10}\,8}=3,\;\Delta_{27\,8}=5,\;\Delta_{27\,10}%
=2,\;\Delta_{35\,10}=6,\;\Delta_{8\,\overline{10}}=-3,\;\Delta_{27\,\overline
{10}\,}=2,\;\Delta_{35\,\overline{10}\,}=6,
\]
and the matrix elements $\delta_{B(R)}^{R^{\prime}}$:%
\begin{equation}%
\begin{tabular}
[c]{ccc}%
$\delta_{N}^{8}=\frac{3}{10},$ & $(\delta_{N}^{\overline{10}})^{2}%
=\frac{1}{20},$ & $\left(  \delta_{N}^{27}\right)  ^{2}=\frac{6}{100},$\\
$\delta_{\Sigma}^{8}=-\frac{1}{10},$ & $(\delta_{\Sigma}^{\overline{10}}%
)^{2}=\frac{1}{20},$ & $\left(  \delta_{\Sigma}^{27}\right)  ^{2}%
=\frac{4}{100},$\\
$\delta_{\Lambda}^{8}=\frac{1}{10},$ & $(\delta_{\Lambda}^{\overline{10}}%
)^{2}=0,$ & $\left(  \delta_{\Lambda}^{27}\right)  ^{2}=\frac{9}{100},$\\
$\delta_{\Xi}^{8}=-\frac{2}{10},$ & $(\delta_{\Xi}^{\overline{10}})^{2}=0,$ &
$\left(  \delta_{\Xi}^{27}\right)  ^{2}=\frac{6}{100},$%
\end{tabular}
\end{equation}
for the octet, and for the decuplet:%
\begin{equation}%
\begin{tabular}
[c]{ccc}%
$\delta_{\Delta}^{10}=\frac{1}{8},$ & $\left(  \delta_{\Delta}^{27}\right)
^{2}=\frac{15}{128},$ & $\left(  \delta_{\Delta}^{35}\right)  ^{2}%
=\frac{25}{896},$\\
$\delta_{\Sigma^{\ast}}^{10}=0,$ & $\left(  \delta_{\Sigma^{\ast}}%
^{27}\right)  ^{2}=\frac{8}{128},$ & $\left(  \delta_{\Sigma^{\ast}}%
^{35}\right)  ^{2}=\frac{40}{896},$\\
$\delta_{\Xi^{\ast}}^{10}=-\frac{1}{8},$ & $\left(  \delta_{\Xi^{\ast}}%
^{27}\right)  ^{2}=\frac{3}{128},$ & $\left(  \delta_{\Xi^{\ast}}^{35}\right)
^{2}=\frac{45}{896},$\\
$\delta_{\Omega}^{10}=-\frac{2}{8},$ & $\left(  \delta_{\Omega}^{27}\right)
^{2}=0,$ & $\left(  \delta_{\Omega}^{35}\right)  ^{2}=\frac{40}{896}.$%
\end{tabular}
\end{equation}
Finally for the antidecuplet we have:%
\begin{equation}%
\begin{tabular}
[c]{cccc}%
$\delta_{\Theta^{+}}^{\overline{10}}=\frac{2}{8},$ & $\left(  \delta
_{\Theta^{+}}^{8}\right)  ^{2}=0,$ & $\left(  \delta_{\Theta^{+}}^{27}\right)
^{2}=0,$ & $(\delta_{\Theta^{+}}^{\overline{35}})^{2}=\frac{72}{896},$\\
$\delta_{\text{N}^{\ast}}^{\overline{10}}=\frac{1}{8},$ & $\left(
\delta_{\text{N}^{\ast}}^{8}\right)  ^{2}=\frac{1}{20},$ & $\left(
\delta_{\text{N}^{\ast}}^{27}\right)  ^{2}=\frac{3}{640},$ & $(\delta
_{\text{N}^{\ast}}^{\overline{35}})^{2}=\frac{81}{896},$\\
$\delta_{\Sigma_{\overline{10}}}^{\overline{10}}=0,$ & $(\delta_{\Sigma
_{\overline{10}}}^{8})^{2}=\frac{1}{20},$ & $(\delta_{\Sigma_{\overline{10}}%
}^{27})^{2}=\frac{8}{640},$ & $(\delta_{\Sigma_{\overline{10}}}^{\overline
{35}})^{2}=\frac{72}{896},$\\
$\delta_{\Xi_{3/2}}^{\overline{10}}=-\frac{1}{8},$ & $(\delta_{\Xi_{3/2}}%
^{8})^{2}=0,$ & $(\delta_{\Xi_{3/2}}^{27})^{2}=\frac{15}{640},$ &
$(\delta_{\Xi_{3/2}}^{\overline{35}})^{2}=\frac{45}{896}.$%
\end{tabular}
\end{equation}

Equations (\ref{H0}), (\ref{2nd}) and (\ref{2nd}) form a complete mass formula
for baryons in all allowed representations which depends upon $4$ free
parameters $M_{\text{cl}}$, $I_{1}$, $I_{2}$ and $\alpha$, or equivalently on%

\begin{equation}
\varepsilon=I_{2}\alpha^{2}%
\end{equation}
$\alpha$, $M_{8}$ and $M_{10}$. By fixing these parameters from the nonexotic
baryonic spectra we can \emph{predict} masses in the exotic antidecuplet. For
example the mass of $\Theta^{+}$ reads:%
\begin{equation}
M_{\Theta^{+}}=M_{8}+\frac{3}{2}\frac{\alpha^{2}}{\varepsilon}-\frac{2}%
{8}\alpha-\frac{3}{112}\varepsilon.
\end{equation}

Similarly to the previous case (\ref{Guad}) the new sum rules can be derived
by examining the null vectors of the $m_{s}$ correction matrix. For example
the following mass relation%
\begin{equation}
\frac{132}{65}\text{N}+\frac{11}{5}\Xi=\frac{33}{26}\Sigma+\frac{77}%
{26}\Lambda\label{SR8}%
\end{equation}
is satisfied with a few promile accuracy. Similar sum rule holds for the
decuplet:%
\begin{equation}
\frac{2627}{594}\Delta+\frac{213}{22}\Xi^{\ast}=\frac{1136}{99}\Sigma^{\ast
}+\frac{71}{27}\Omega.\label{SR10}%
\end{equation}
Somewhat awkward coefficients in Eqs.(\ref{SR8},\ref{SR10}) are due to the
SU(3) Clebsch--Gordan coefficients \cite{DeSw} entering formula (\ref{2nd}).

\begin{flushleft}
{\bf 4. Numerical estimates}
\end{flushleft}

Question arises how to use the mass formula given as a sum of $H_{0}$ of
Eq.(\ref{H0}) $M_{B(R)}^{(1)}$ and $M_{B(R)}^{(2)}$ given by Eqs.(\ref{1st}%
,\ref{2nd}). In our early work from 1987 \cite{Mogtalk} we have estimated the
value of parameter $\alpha$ and then fitted the remaining 3 parameters
$M_{8,10}$ and $\varepsilon$ by minimizing the functional%
\begin{equation}
F[M_{8,}M_{10,}\alpha,\varepsilon]=\frac{1}{18}\sum\limits_{B}\left(
M_{B}^{\text{th}}[M_{8,}M_{10,}\alpha,\varepsilon]-M_{B}^{\text{exp}}\right)
^{2}\label{Func}%
\end{equation}
with $\alpha$ fixed. In 1987 the accepted value of the pion-nucleon sigma term
was at the level of $60$ MeV \cite{sigma} in contrast to the present value of
$45$ MeV \cite{sigma1} used in Ref.\cite{DPP}. The average light quark mass
was taken as $\overline{m}=5.5$ MeV which gave $\sigma=3.64$. Taking
$m_{s}\simeq200$ MeV we got $\alpha\simeq720$ MeV. By minimizing (\ref{Func})
with respect to the remaining parameters we got the $\Theta^{+}$ mass equal to
$1534$ MeV. This is fit no. I. In fit no. II we have allowed all parameters to
be free with the result $M_{\Theta^{+}}=1339$ MeV. The parameters for the fits
are given in Table \ref{tab:fits}.

\begin{table}[ptb]
\caption{Parameters of the model and corresponding moments of inertia in MeV}%
\label{tab:fits}
\begin{center}%
\begin{tabular}
[c]{ccccccc}
& $M_{8}$ & $M_{10}$ & $\alpha$ & $\varepsilon$ & $1/I_{1}$ & $1/I_{2}$\\
\text{fit I} & 1213 & 1503 & 720 & 1442 & 193 & 360\\
\text{fit II} & 1230 & 1535 & 616 & 1824 & 205 & 284
\end{tabular}
\end{center}
\end{table}

\begin{figure}[ptbh]
\begin{center}
\includegraphics[scale=1.0]{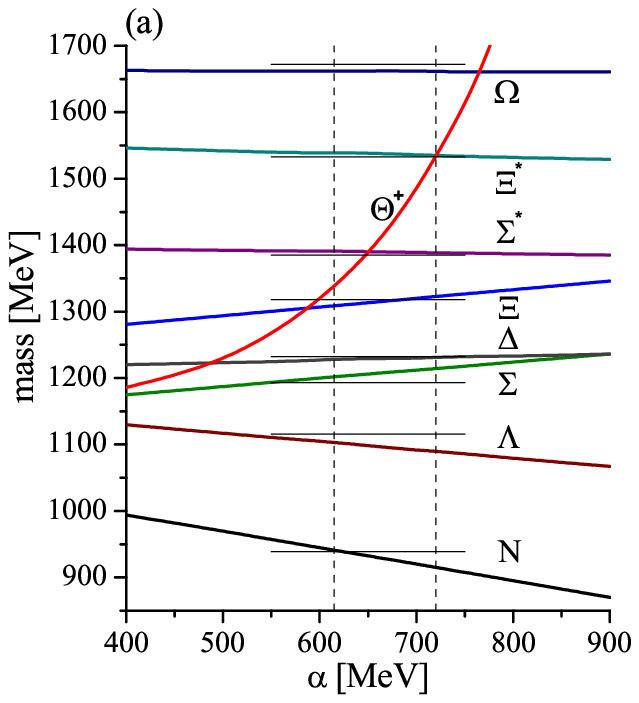}  \hspace{0.3cm}
\includegraphics[scale=1.0]{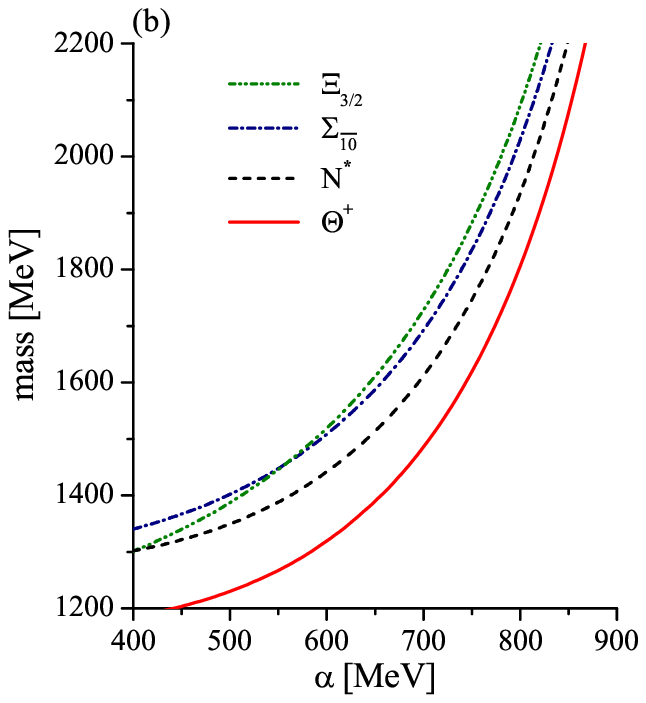}
\end{center}
\caption{(a) Spectrum of the nonexotic baryons as a function of parameter
$\alpha$, together with a prediction for the mass of $\Theta^{+}$. Thin
horizontal lines correspond to the experimental data, dashed vertical lines
correspond to fit II ($\alpha=615$) and fit I ($\alpha=720$).~ (b) Spectrum of
the exotic antidecuplet as a function of parameter $\alpha$. }%
\label{fig:f1}%
\end{figure}

Unfortunately the mass of $\Theta^{+}$ is quite sensitive to the choice of
parameters. This is illustrated in Fig.\ref{fig:f1}.a where we plot the
results of the constrained fits obtained by minimizing (\ref{Func}) for fixed
$\alpha$. We see a rather steep rise of $M_{\Theta^{+}}$ with $\alpha$. Two
dashed vertical lines correspond to $\alpha$ of fits I and II and thin
horizontal lines to the experimental masses. 

Although there is rather large sensitivity of $M_{\Theta^{+}}$ to the choice
of $\alpha$, the conclusion one can draw from Fig.\ref{fig:f1}.a is that
$\Theta^{+}$ is relatively light. The full spectrum of the exotic antidecuplet
as a function of parameter $\alpha$ is shown in Fig.\ref{fig:f1}.b We see that
below $\alpha=600$ MeV the second order correction reshuffles the order of the
antidecuplet states which means that the perturbation theory becomes
unreliable. As compared to the first order perturbation theory, the states are
no longer equally spaced, with $\Theta^{+}$ being significantly lighter than
the other members of $\overline{10}$. It is interesting to note that if we
fixed the value of $I_{2}$ (or equivalently $\alpha$) by the requirement that
$M_{\text{N}^{\ast}}=1710$ MeV, we would get $\alpha=738$ MeV corresponding to
$\Theta^{+}$ mass of $1580$ MeV.

Finally let us present another estimate of $I_{2}$ based on the model formulae
for the parameters of Hamiltonian $H_{0}$ in the \emph{first} order of the
perturbation theory. These parameters depend on the soliton size, $r_{0}$,
which enters the convenient variational Ansatz \cite{arcDP} of the soliton
profile function $P(r)$ (\ref{hedgehog}):%
\begin{equation}
P(r)=2\arctan\left[  \left(  \frac{r_{0}}{r}\right)  ^{2}\right]
.\label{arctan}%
\end{equation}
In practice one uses dimensionless variable $x_{0}=eF_{\pi}r_{0}$. Here $e$ is
a free parameter corresponding to the Skyrme term in the mesonic Lagrangian
(\ref{Lagr}) and $F_{\pi}=186$ MeV (for the details see \emph{e.g.}
Ref.\cite{MPh}). In Ref.\cite{MPh} we have calculated $M_{\text{cl}}$, and
$I_{1,2}$ in terms of $x_{0}$ for the arctan Ansatz (\ref{arctan}):
\begin{align}
M_{\mathrm{cl}}  &  =\frac{F_{\pi}}{e}{\pi}^{2}\frac{3\sqrt{2}}{16}%
(4x_{0}+\frac{15}{x_{0}}),\label{Mcl}\\
I_{1}  &  =\frac{1}{e^{3}F_{\pi}}\pi^{2}\frac{\sqrt{2}}{12}\,(6x_{0}%
^{3}+25x_{0}),\label{I1}\\
I_{2}  &  =\frac{1}{e^{3}F_{\pi}}\pi^{2}\frac{\sqrt{2}}{16}\,(4x_{0}%
^{3}+9x_{0}).\label{I2}%
\end{align}
By minimizing $M_{\text{cl}}$ with respect to the soliton size we get we get
$x_{0}=\sqrt{15/4}$. It is easy to convince oneself that by fitting the
octet--decuplet mass difference:%
\begin{equation}
M_{10}-M_{8}=\frac{3}{2I_{1}}=231\;\text{MeV}%
\end{equation}
we get $e=4.64$. With this value of $e$ we have $1/I_{2}\simeq400$ MeV and
$M_{\overline{10}}=1762\;$MeV. Assuming that splittings in $\overline{10}$ are
identical with the splittings in the ordinary decuplet (which is true in the
\emph{first} order of the perturbation theory for the breaking Hamiltonian
(\ref{HSM})) we get $M_{\Theta^{+}}\simeq1460$ MeV.

\begin{flushleft}%
{\bf 5. Comparison with the Chiral Quark Model}%

\end{flushleft}

The authors of Ref.\cite{DPP} used the mass formula in the first order in
$m_{s}$ with, however, nonleading terms in $N_{c}$ (see Eq.(\ref{HbrQM})). The
reason was, as in our case, that the mass formula in the leading order in
$N_{c}$ and in $m_{s}$ (\ref{Hprim}) was unable to reproduce the nonexotic
mass spectra with good enough accuracy. Systematic expansion of the rotational
effective Hamiltonian in inverse powers of $N_{c}$ \cite{Blotz} introduces $3$
unknown free constants $\alpha^{\prime}$, $\beta$ and $\gamma$ to $H^{\prime}$
of Eq.(\ref{HbrQM}). However, only two linear combinations enter the nonexotic
mass splittings. Therefore the ordinary baryons allow to fix only $4$
parameters $M_{8}$, $M_{10}$ and the above mentioned two linear combinations
of $\alpha^{\prime}$, $\beta$ and $\gamma$. In other words, in contrast to our
approach, there is no way to fix $I_{2}$ without some further assumptions.
Also the third linear combination of breaking parameters cannot be fixed. This
is an important ingredient, because the exotic spectra are sensitive to other
linear combinations of $\alpha^{\prime}$, $\beta$ and $\gamma$ \cite{DPP}.

\begin{figure}[ptbh]
\begin{center}
\includegraphics[scale=1.0]{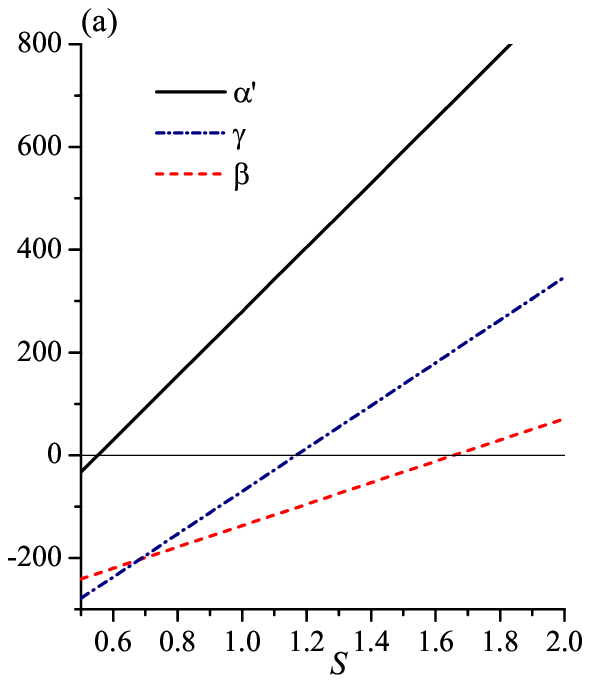}  \hspace{0.3cm}
\includegraphics[scale=1.0]{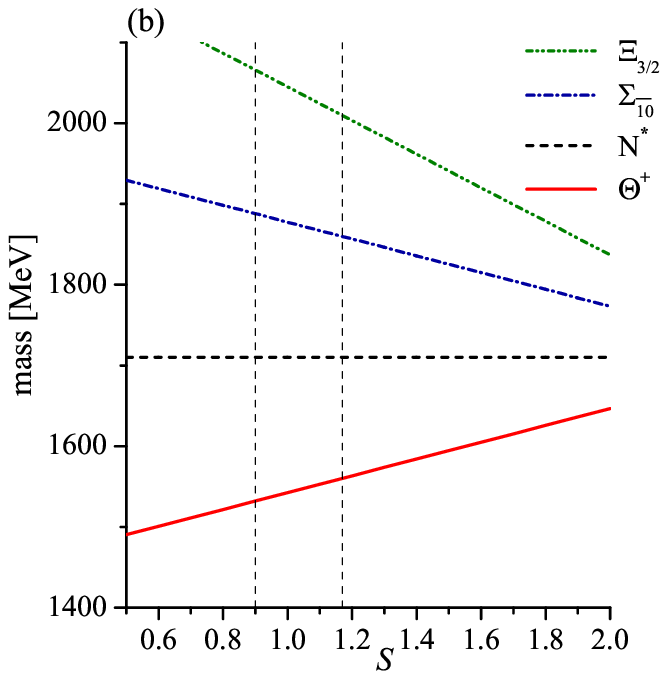}
\end{center}
\caption{(a) Parameters \ $\alpha^{\prime},\beta$ and \ $\gamma$ as functions
of \ $S$ constrained to fit the nonexotic baryon spectra.~ (b) Spectrum of
antidecuplet in the Chiral Quark Model as a function of $S$. $S=0.9$
corresponds to the choice of Ref.[13], whereas $S=1.16$ corresponds to
$\gamma=0$. }%
\label{fig:f2}%
\end{figure}

The main uncertainty due to $I_{2}$ was removed in Ref.\cite{DPP} by the
assumption that the ''nucleonic'' member of the antidecuplet was identified
with the nucleon resonance N$^{\ast}$(1710). The remaining freedom,
illustrated in Fig.\ref{fig:f2}.a was removed by fixing $m_{s}=150$ MeV and
$\Sigma_{\pi N}=45$ MeV. How big is the residual uncertainty due to the choice
of $m_{s}$ and $\Sigma_{\pi N}$? We illustrate this in Fig.\ref{fig:f2}.b
where we plot the spectrum of the antidecuplet states, for fixed N$^{\ast}%
$(1710), as a function of a dimensionless parameter%
\begin{equation}
S=\frac{m_{s}}{150}\frac{\Sigma_{\pi N}}{50}.\label{Sdef}%
\end{equation}
The choice of Ref.\cite{DPP} corresponds to $S=0.9$. However, other choices of
$S$ are phenomenologically not excluded. For example in Refs.\cite{gam0} we
have used parameterization with $\gamma=0$, as the explicit model calculations
suggest \cite{Blotz}, which corresponds to $S=1.16$. With this choice of $S$
we get $M_{\Theta^{+}}=1560$ MeV. This is very close to $1580$ MeV obtained
within the Skyrme model second order perturbation theory by fixing N$^{\ast}$
at $1710$ MeV.

Our original choice of $m_{s}$ and $\Sigma_{\pi N}$ \cite{Mogtalk} corresponds
to $S=1.6$, \emph{i.e.} as seen from Fig.\ref{fig:f2}.a, to $\beta=0$.
Although in our approach $\gamma=0$ as well, however, the second order
perturbation theory generates effects similar to $\gamma\ne0$. Figure
\ref{fig:f2}.a indicates that the good fit to the nonexotic spectra with
$S=1.6$ requires rather large $\alpha\simeq700$ MeV in agreement with our fit I.

\begin{flushleft}
{\bf 6. Discussion and conclusions}
\end{flushleft}

The above discussion shows that chiral models predict the existence of the low
lying exotic baryons. The precise value of the lightest member of the exotic
antidecuplet depends crucially on the estimate of the strange moment of
inertia $I_{2}$. We have presented two kinds of estimates of $I_{2}$. The one
based on the second order mass formula for baryons, allowed us to fix $I_{2}$
from the spectrum of the nonexotic states . We used the second order
perturbation theory in order to fit the nonexotic spectra with high accuracy.
With our original choice of parameters for the pion-nucleon sigma term
$\Sigma_{\pi N}=60$ MeV and the strange quark mass $m_{s}=200$ MeV we have
predicted $M_{\Theta^{+}}=1534$ MeV. This number is in striking agreement with
the latest experimental evidence, however, one should not forget that in fact
the prediction of $M_{\Theta^{+}}$ depends rather strongly on the model
parameters. If we used, as in Ref.\cite{DPP}, the mass of N$^{\ast}(1710)$ to
anchor the center of $\overline{10}$, we would get $M_{\Theta^{+}}=1560$. A
similar analysis has been recently done in Ref.\cite{KopWal}.

The second estimate of $I_{2}$ was based on the first order mass formula and
on model expression for $I_{2}$ (\ref{I2}). With this method the mass of
$\Theta^{+}$ comes out slightly lower; $M_{\Theta^{+}}\simeq1460$.

The existence of the light, exotic, positive parity baryons belonging to
antidecuplet is a natural feature of the chiral models \cite{Chem}%
,\cite{MatKar}\cite{KarMog},\cite{Mogtalk}. From the beginning it was clear
that the minimal quark content of such a multiplet must be $q^{4}\overline{q}$
\cite{KarMog}. Quark models, however, have difficulties in accommodating
$\Theta^{+}$ of \emph{positive} parity (yet requiring experimental
confirmation) unless strong interquark correlations are introduced
\cite{JafWil},\cite{KarLip}. Perhaps the most striking difference between the
quark and the soliton pictures consists in the fact that in the quark models
$\overline{10}$ is inevitably accompanied by an exotic octet which does not
appear in the soliton approach. With the discovery of $\Theta^{+}$ low energy
QCD regained attention and the constituent quark \emph{vs.} soliton
interpretations of the light baryons compete in explaining the data.

We would like to thank B. Jaffe, M. Polyakov, V. Petrov and K. Goeke for
conversations. We benefited from the e-mail exchanges with K. Hicks, D.I.
Diakonov, M. Karliner and H.J. Lipkin. This work was partly supported by the
Polish State Committee for Scientific Research under grant 2 P03B 043 24.


\begin{thebibliography}{9}                                                                                                %
\bibitem {nakano}LEPS Collaboration (T. Nakano \emph{et al.}), Phys. Rev.
Lett. \textbf{91} (2003) 012002.

\bibitem {Diana}DIANA Collaboration (V.V. Barmin \emph{et al.}), hep-ex/0304040.

\bibitem {CLAS}CLAS Collaboration (S. Stepanyan \emph{et al.}), hep-ex/0307018.

\bibitem {SAPHIR}SAPHIR Collaboration (J. Barth \emph{et al.}), hep-ex/0307083.

\bibitem {man}A. Manohar, Nucl. Phys \textbf{B248} (1984) 19.

\bibitem {Chem}M. Chemtob, Nucl. Phys. \textbf{B256} (1985) 600.

\bibitem {Gua}E. Guadagnini, Nucl. Phys. \textbf{B236} (1984) 35.

\bibitem {NMP}P.O. Mazur, M. Nowak and M. Prasza\l owicz, Phys. Lett.
\textbf{B147} (1984) 137.

\bibitem {JW}S. Jain and S.R. Wadia, Nucl. Phys. \textbf{B258} (1985) 713.

\bibitem {MatKar}M.P. Mattis and M. Karliner, Phys. Rev. \textbf{D34} (1986) 1991.

\bibitem {KarMog}M. Karliner talk at \emph{Workshop on Skyrmions and
Anomalies}, M. Je\.{z}abek and M. Prasza\l owicz editors, World Scientific
1987, page 164.

\bibitem {Mogtalk}M. Prasza{\l}owicz, talk at \emph{Workshop on Skyrmions and
Anomalies}, M. Je\.{z}abek and M. Prasza{\l}owicz editors, World Scientific
1987, page 112.

\bibitem {DPP}D.I. Diakonov, V.Yu. Petrov and M.V. Polyakov, Z. Phys.
\textbf{A359} (1997) 305.

\bibitem {Drev}For a review see D.I. Diakonov and V.Yu. Petrov, contribution
to the Festschrift in honor of B.L. Ioffe, M. Shifman editor, \emph{At the
frontier of particle physics}, vol. 1, page 359, hep-ph/0009006; D.I.
Diakonov, lectures given at Advanced Summer School on Nonperturbative Quantum
Field Physics, Peniscola, Spain, 2-6 June 1997, \emph{Advanced school on
non-perturbative quantum field physics} 1-55, hep-ph/9802298.

\bibitem {NJL}Y. Nambu and G. Jona-Lasinio, Phys. Rev. \textbf{124} (1961) 246.

\bibitem {instvac}D.I. Dyakonov and V.Yu. Petrov, Nucl. Phys. \textbf{B245}
(1984) 259; \textbf{B272} (1986) 457; E.V. Shuryak, Phys. Rep. \textbf{115}
(1985) 15.

\bibitem {grad}D.I. Dyakonov and M. Eides JETP Lett. \textbf{38} (1983) 433;
J. \.{Z}uk, Z. Phys. \textbf{29} (1985) 303; M. Prasza\l owicz and G.
Valencia, Nucl. Phys. \textbf{B341} (1990) 27; E. Ruiz Arriola Phys. Lett.
\textbf{B253} (1991) 430.

\bibitem {Skyrme}T.H.R Skyrme, Proc. Royal Soc. \textbf{A260} (1961) 127;
Nucl. Phys. \textbf{31} (1962) 556.

\bibitem {WZ}J. Wess and B. Zumino, Phys. Lett. \textbf{B37} (1971) 95.

\bibitem {Witten}E. Witten, Nucl. Phys. \textbf{B160} (1979) 57; \textbf{B223}
(1983) 422; \textbf{B223} (1983) 433.

\bibitem {ANW}G.S. Adkins, C.R. Nappi and E. Witten, Nucl. Phys. \textbf{B228}
(1983) 552; G.S. Adkins and C.R. Nappi, Nucl.Phys. \textbf{B233} (1984) 109.

\bibitem {MPSU3}M. Prasza\l owicz, Phys. Lett. \textbf{B158} (1985) 264.

\bibitem {Weigel}For an extensive review of the SU(3) Skyrme Model see H.
Weigel, Int. Jour. of. Mod. Phys., \textbf{11} (1996) 2419.

\bibitem {AN}G.S. Adkins and C.R. Nappi, Nucl. Phys. \textbf{B249} (1985) 507.

\bibitem {Cas}B. Moussallam and D. Kalafatis, Phys. Lett. \textbf{B272} (1991)
196; B. Moussallam, Annals Phys. \textbf{225} (1993) 264.

\bibitem {Blotz}A. Blotz, D.I. Diakonov, K. Goeke, N.W. Park, V.Yu. Petrov,
P.V. Pobylitsa, Phys. Lett. \textbf{B287} (1992) 29; Nucl. Phys. \textbf{A555}
(1993) 765; H. Weigel, R. Alkofer and H. Reinhardt, Phys. Lett. \textbf{B284}
(1992) 296.

\bibitem {DeSw}J.J. De Swart, J. of Mod. Phys. \textbf{35} (1963) 916.

\bibitem {YA}H. Yabu and K. Ando, Nucl. Phys. \textbf{B301} (1988) 601.

\bibitem {MPgA}M. Praszalowicz, Phys. Rev. \textbf{D42} (1990) 216; P. Sieber,
M. Praszalowicz and K. Goeke, Nucl. Phys. \textbf{A569} (1994) 629.

\bibitem {sigma}R. Koch, Z. Phys. \textbf{C15} (1982) 161; W. Wiedner \emph{et
al}., Phys. Rev. Lett. \textbf{58} (1987) 648; J. Gasser, H. Leutwyler, M.P.
Locher and M.E. Sainio, Phys. Lett. \textbf{B213} (1988) 85.

\bibitem {sigma1}J. Gasser, H. Leutwyler and M.E. Sainio, Phys. Lett.
\textbf{B253} (1991) 252.

\bibitem {arcDP}D.I. Dyakonov and V.Yu. Petrov, JETP Lett. \textbf{43} (1986) 57.

\bibitem {MPh}M. Prasza\l owicz, Acta Phys. Pol. \textbf{B22} (1991) 525.

\bibitem {gam0}H-Ch. Kim, M. Prasza\l owicz and Klaus Goeke, Phys. Rev.
\textbf{D57} (1998) 2859, H-Ch. Kim, M. Prasza\l owicz, M.V. Polyakov and
Klaus Goeke, Phys. Rev. \textbf{D58} (1998) 114027.

\bibitem {KopWal}V.B. Kopeliovich and H. Walliser, hep-ph/0304058.

\bibitem {JafWil}R.L. Jaffe and F. Wilczek, hep-ph/0307341.

\bibitem {KarLip}M. Karliner and H.J. Lipkin, hep-ph/0307243.
\end{thebibliography}
\end{document}